\documentclass[prm,aps,twocolumn,showpacs,superscriptaddress]{revtex4-1}
\usepackage{hyperref}
\hypersetup{colorlinks=true, citecolor=blue, filecolor= black, linkcolor= blue, urlcolor= blue}
\usepackage[utf8]{inputenc}
\usepackage{graphicx}
\usepackage{dcolumn}
\usepackage{bm}
\usepackage{float}
\usepackage{gensymb}
\usepackage{braket}
\usepackage{bm}
\usepackage[namelimits]{amsmath} %
\usepackage{amsfonts}            %
\usepackage{mathrsfs}            %

\begin{document}


\title{Negative linear compressibility and unusual dynamic behaviors of NaB$_{3}$}

\author{Xin-Ling He}
\affiliation{Key Laboratory of Weak-Light Nonlinear Photonics and School of Physics, Nankai University, Tianjin 300071, China}

\author{Shu-Ning Pan}
\affiliation{National Laboratory of Solid State Microstructures, School of Physics and Collaborative Innovation Center of Advanced Microstructures, Nanjing University, Nanjing 210093, China}

\author{Yue Chen}
\affiliation{Department of Mechanical Engineering, The University of Hong Kong, Pokfulam Road, Hong Kong SAR, China}
\affiliation{HKU Zhejiang Institute of Research and Innovation, 1623 Dayuan Road, Lin An 311305, China}

\author{Xiao-Ji Weng}
\affiliation{Key Laboratory of Weak-Light Nonlinear Photonics and School of Physics, Nankai University, Tianjin 300071, China}

\author{Zifan Wang}
\affiliation{Key Laboratory of Weak-Light Nonlinear Photonics and School of Physics, Nankai University, Tianjin 300071, China}

\author{Dongli Yu}
\affiliation{Center for High Pressure Science, State Key Laboratory of Metastable Materials Science and Technology, School of Science, Yanshan University, Qinhuangdao 066004, China}

\author{Xiao Dong}
\email{xiao.dong@nankai.edu.cn}
\affiliation{Key Laboratory of Weak-Light Nonlinear Photonics and School of Physics, Nankai University, Tianjin 300071, China}

\author{Jian Sun}
\email{jiansun@nju.edu.cn}
\affiliation{National Laboratory of Solid State Microstructures, School of Physics and Collaborative Innovation Center of Advanced Microstructures, Nanjing University, Nanjing 210093, China}

\author{Yongjun Tian}
\affiliation{Center for High Pressure Science, State Key Laboratory of Metastable Materials Science and Technology, School of Science, Yanshan University, Qinhuangdao 066004, China}

\author{Xiang-Feng Zhou}
\email{xfzhou@nankai.edu.cn}
\email{zxf888@163.com}
\affiliation{Key Laboratory of Weak-Light Nonlinear Photonics and School of Physics, Nankai University, Tianjin 300071, China}
\affiliation{Center for High Pressure Science, State Key Laboratory of Metastable Materials Science and Technology, School of Science, Yanshan University, Qinhuangdao 066004, China}

\begin{abstract}
\noindent First-principles calculations reveal that sodium boride (NaB$_{3}$) undergoes a phase transition from a tetragonal $P$4/$mbm$ phase to an orthorhombic $Pbam$ phase at $\sim$16 GPa, accompanied by counterintuitive lattice expansion along the crystallographic $a$-axis. This unusual compression behavior is identified as negative linear compressibility (NLC), which is dominantly attributed to the symmetry-breaking of boron framework. Meanwhile, the $P$4/$mbm$ and $Pbam$ phases form superionic conductors after undergoing a peculiar swap state at high temperature. Specifically, under ``warm" conditions the Na cation pairs exhibit a rare local exchange (or rotation) behavior, which may be originated from the asymmetric energy barriers of different diffusion paths. The study of NaB$_{3}$ compound sheds new light on a material with the combination of NLC and ion transportation at extreme conditions.
\end{abstract}



\maketitle
 \section{INTRODUCTION}
 High pressure (HP) and high temperature (HT) can significantly alter both crystal structure and electronic properties of a material without creating any external impurities, leading to a tremendous amount of discovery of new compounds and unusual properties. For instance, most matters contract in all directions under hydrostatic pressure, whereas very few materials expand along one direction coupled to the volume reduction due to its anisotropy, which is defined as negative linear compressibility (NLC) \cite{R01,R02,R03}. Several NLC materials were previously reported in dense inorganic oxides and fluorides, as well as a few complex organics \cite{R04,R05,R06,R07,R08,R09,R10,R11}. On the basis of this anomalous mechanical property, the NLC materials have important applications in the design of pressure sensors, actuators, artificial muscles and optical fibers with high shock resistance \cite{R02,R12,R13,R14,R15,R16}. Another example relates to superionic compounds, also denoted as fast ion conductors or solid electrolytes, whose structures usually contain the rigid frameworks with open channels along which ions can migrate \cite{R17,R18,R19}. Superionic materials featured by a liquid-like conductivity within the fixed crystalline frameworks have long attracted enormous attention for battery, fuel cell, thermoelectrics, and other energy applications, e.g., Li$_{10}$GeP$_{2}$S$_{12}$, LiAlSO and Cu$_{2}$Se \cite{R20,R21,R22}. Furthermore, a few common materials such as ice and ammonia exhibit superionic state at HPHT conditions, which even improves our knowledge of the middle ice layers of Neptune and Uranus \cite{R17,R18}. For sodium borides, only a few compounds are precisely determined with regard to their compositions and structures \cite{R23,R24,R25}. The new predicted phase of $I2_{1}2_{1}2_{1}$-Na$_{2}$B$_{30}$ with an unprecedented open-framework boron sublattice is helpful to resolve the debate on the ground-state structure of sodium borides \cite{R23,R24,R25,R26}. Those results inspired us to further explore the phase diagram and properties of sodium borides at mild pressure. In this work, we surprisingly found sodium boride (NaB$_{3}$) had a temperature-induced swap state beside superionicity and NLC simultaneously, providing a representative example to better understand material's structure and properties in general.

\section{METHOD}
Structure searches were performed utilizing the \textit{ab initio} evolutionary algorithm USPEX \cite{R27,R28} for NaB$_{3}$ with up to six formula units (f.u.) per supercell under various pressures. The structure relaxations and electronic properties were carried out using density functional theory (DFT) within the Perdew-Burke-Ernzerhof (PBE) parametrization of the generalized gradient approximation (GGA) exchange-correlation functional \cite{R29}, as implemented in the VASP code \cite{R30} with 2$s^{2}$2$p^{6}$3$s^{1}$ and 2$s^{2}$2$p^{1}$ treated as valence electrons for Na and B atoms, respectively. A plane-wave basis set with an energy cutoff of 850 eV and uniform $\Gamma$-centered $k$-point grids with a resolution of $2 \pi \times 0.02$~\AA$^{-1}$ were used in the electronic self-consistent calculations. The structures were fully optimized until the maximum energy and force were less than 10$^{-8}$ eV and 0.001 eV/{\AA}, respectively. Electronic properties were also calculated by using the Heyd-Scuseria-Ernzerhof (HSE06) functional \cite{R31}. \textit{Ab initio} molecular dynamics (AIMD) simulations \cite{R32} were carried out using $2 \times 2 \times 3$ supercell for the $P$4/$mbm$ and $Pbam$ phases, and $2 \times 2 \times 2$ supercell for the $I$-$4m2$ phase. A Nos\'e-Hoover thermostat was adopted to perform the $NVT$ simulations with the temperature range of 500-4500 K lasting for 10 ps with a time step of 1 fs. The statistical information of each atomic trajectory was extracted from the last 5 ps, and some trajectories were extended to 40 ps to confirm the stability of the simulations. Phonon dispersion curve of $P$4/$mbm$ NaB$_{3}$ was performed using the finite displacement method ($2 \times 2 \times 2$ supercell) with the \textsc{phonopy} code \cite{R33}. In addition, the energy barrier was calculated by the climbing-image nudged elastic band (CI-NEB) method \cite{R34}.

\begin{figure}[t]
\begin{center}
\includegraphics[width=8.0cm]{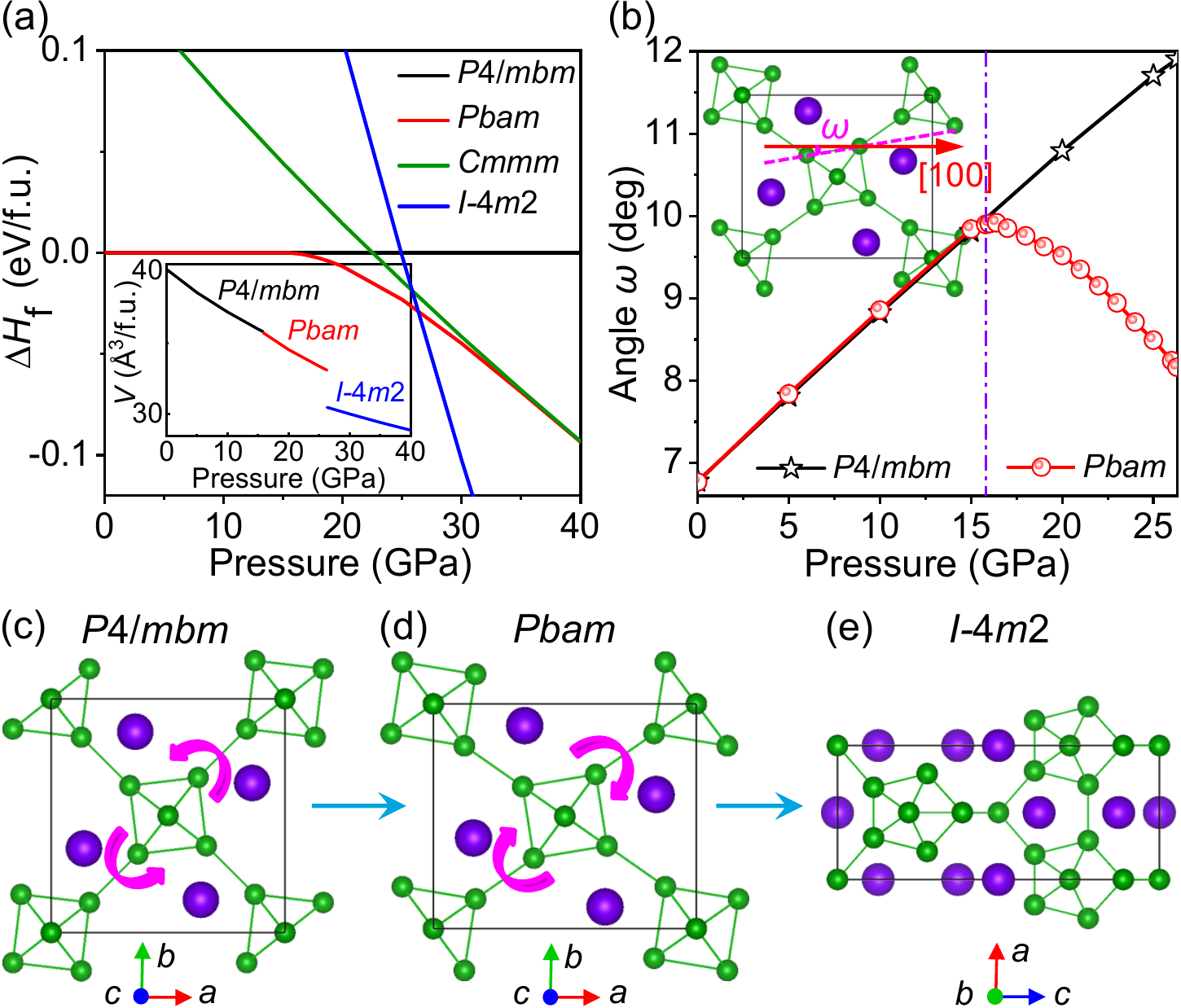}
\caption{%
(a) The enthalpy difference of NaB$_{3}$ compounds as a function of pressure with respect to $P$4/$mbm$ phase. The insert shows the pressure-volume relations. (b) The variation of the angle $\omega$ as function of pressure. The inset shows the angle $\omega$ between the orientation of the centered B$_{6}$ octahedral and [100] direction. (c-e) Schematic views of structural evolutions of NaB$_{3}$ system during compression. The rotation direction of B$_{6}$ octahedral is labeled by magenta arrows as increasing pressure. Purple and green balls represent the Na and B atoms, respectively.}
\end{center}
\end{figure}

\begin{figure}[t]
\begin{center}
\includegraphics[width=8.0cm]{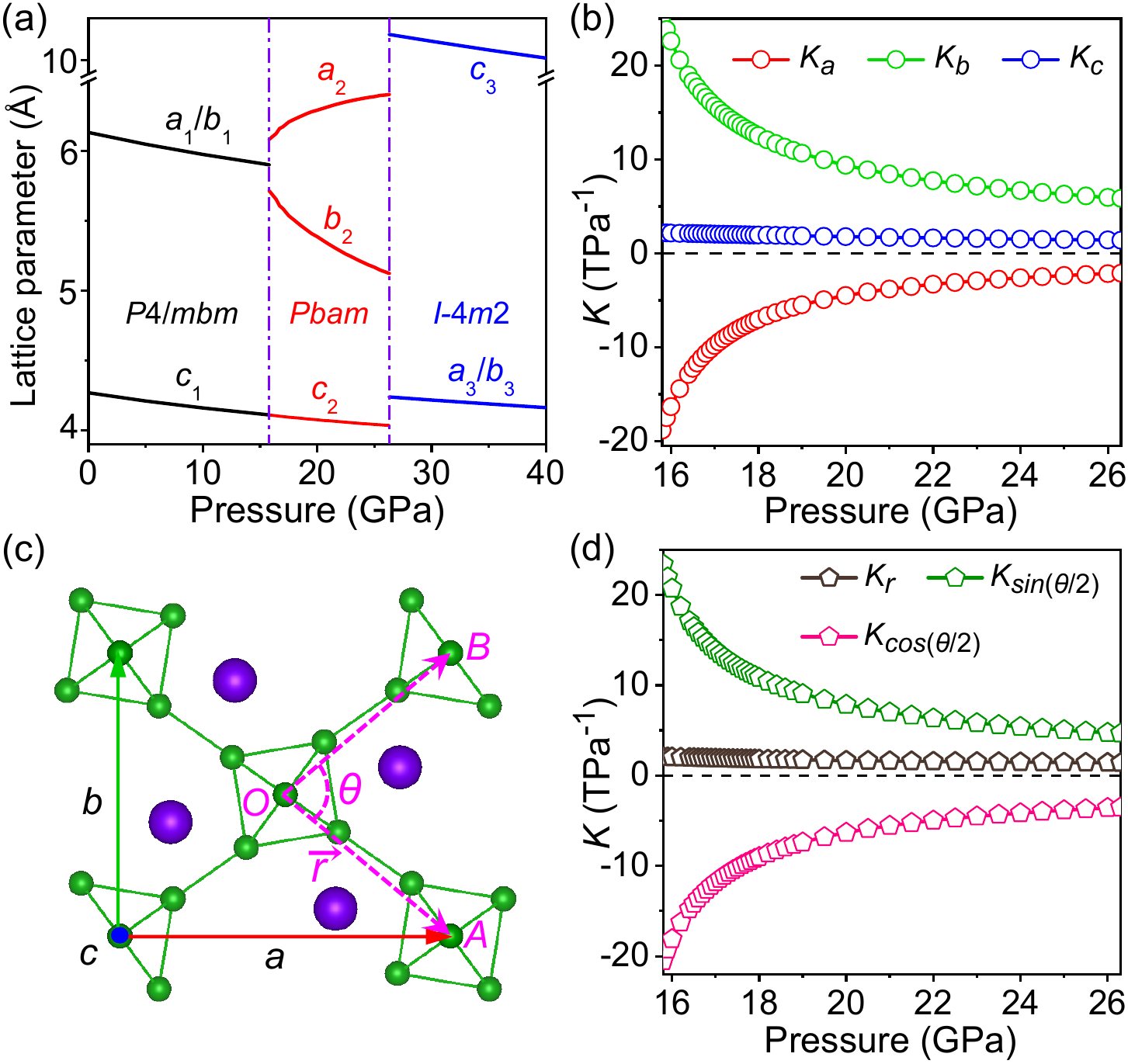}
\caption{%
(a) The lattice parameters for the predicted stable phases as a function of pressure. (b) The compressibility of the principal axes for $Pbam$ phase as a function of pressure. (c) The structural projection of $Pbam$ phase along the [001] direction, where the letters $A$, $O$ and $B$ are the centers of B$_{6}$ octahedra. The pressure-independent coordinates for $A$, $O$ and $B$ are (1, 0, 1/2), (1/2, 1/2, 1/2) and (1, 1, 1/2), respectively. (d) Evolution of compressibility for the $O$-$A$ or $O$-$B$ distance $r$, $\sin$($\theta$/2) and $\cos$($\theta$/2) within the pressure range of 15.8-26.3 GPa.}
\end{center}
\end{figure}

\section{RESULTS AND DISCUSSION}
The phase diagram and physical properties of NaB$_{3}$ are systematically investigated. Since the enthalpy of formation of $P$4/$mbm$-NaB$_{3}$ is negative (Fig. S1) referring to the enthalpies of $bcc$-Na and $\alpha$-boron structure or any isochemical mixture of Na-B compounds \cite{R35}, first-principles calculations reveal that $P$4/$mbm$-NaB$_{3}$ system is stable above $\sim$0.4 GPa and may be quenchable to ambient pressure because there are no imaginary phonon frequencies in the whole Brillouin zone (Fig.S2) \cite{R35}. The enthalpy-pressure curves show that NaB$_{3}$ undergoes a reversible phase transition from a tetragonal $P$4/$mbm$ phase to an orthorhombic $Pbam$ phase at $\sim$16 GPa. Note that the $Pbam$ symmetry is the subgroup of $P$4/$mbm$ symmetry with the index of two, and the fourfold rotation symmetry is broken (mainly attributed to the rotation of B$_{6}$ octahedron) as pressure is increased. Moreover, no chemical bonds break during the phase transition, leading to the gradual variation of volume and the reversible phase transition. Then $P$4/$mbm$-NaB$_{3}$ undergoes an irreversible phase transition to another tetragonal $I$-$4m2$ phase at $\sim$26 GPa [Fig. 1(a)]. The phase transition from the $P$4/$mbm$ to $Pbam$ structure is accompanied by the negligible volume change, indicative of the second-order phase transition, while 7.87$\%$ volume collapse for the phase transition from $Pbam$ to $I$-$4m2$ structure, the characteristic of first-order phase transition [the inset of Fig. 1(a)]. These results are different from the reported phase transition from $P$4/$mbm$ to $Cmmm$ structure \cite{R36,R37}, and the $Cmmm$ structure is higher in enthalpy than the newly proposed phases at the studied pressure range.

\begin{figure*}[t]
\begin{center}
\includegraphics[width=13cm]{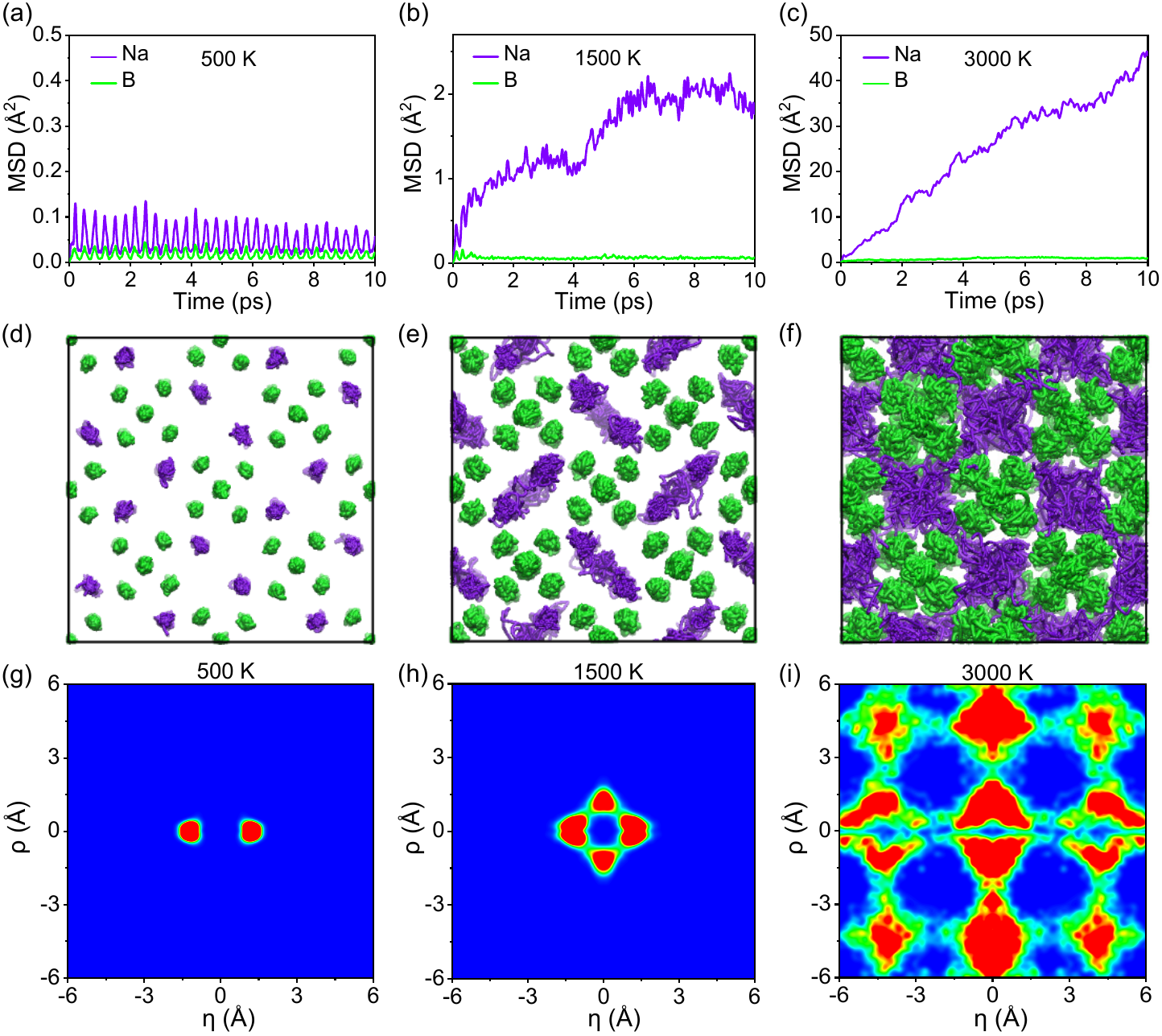}
\caption{%
(a-c) The MSDs from AIMD simulations for the Na and B atoms in $P$4/$mbm$ phase at ambient pressure and different temperatures. (d-f) The projection of atomic trajectories along the [001] direction in $P$4/$mbm$ phase from the last 5 ps run representing the solid phase (500 K), the swap phase (1500 K), and the superionic phase (3000 K). (g-i) The probability of the presence of Na atoms at 500, 1500 and 3000 K.}
\end{center}
\end{figure*}

The crystal structures of the $P$4/$mbm$ and $Pbam$ phases are plotted in Figs. 1(c) and 1(d) for contrast. They are composed of interstitial Na atoms and different boron frameworks, that is, B$_{6}$-octahedral network for $P$4/$mbm$ and $Pbam$ phases whereas the dominant B$_{8}$-dodecahedral network for $I$-$4m2$ phase [Fig. 1(e)]. The angle ($\omega$) between the orientation of the centered B$_{6}$ octahedra (marked by the magenta dotted line) and the [100] direction is represented to explore the mechanism for the reversible phase transition [the inset of Fig. 1(b)]. As shown in Fig. 1(b), the angle $\omega$ in $P$4/$mbm$ phase is linearly increased as a function of pressure owing to the counterclockwise rotation of the centered B$_{6}$ octahedron. In contrast, it abruptly decreases in parabolic-shape from 15.8 GPa to 26.3 GPa because of the opposite rotation behavior of the B$_{6}$ octahedron in $Pbam$ phase. As further increasing pressure above 26 GPa, the B$_{6}$ octahedra are destroyed and transform into distorted B$_{8}$ dodecahedra and interstitial B atoms, resulting in the formation of $I$-$4m2$ phase. The lattice parameters of NaB$_{3}$ under selected pressures are listed in Table SI \cite{R35}. Moreover, band structure calculations show that all predicted NaB$_{3}$ compounds are semiconductors. In contrast, the DFT band gap of $Pbam$ phase is increased as a function of pressure and different from other phases (Fig. S3) \cite{R35}, which is partially originated from the NLC effect.

Unexpectedly, the phase transition from the $P$4/$mbm$ to $Pbam$ phase is quite different from those second-order phase transitions \cite{R38,R39,R40,R41}, because it is accompanied by the emergence of anomalous lattice expansion along the crystallographic $a$-axis (belong to the proper ferroelastic transitions \cite{R05}) whereas normal lattice contractions along the $b$- and $c$-axes within pressure range of $\sim$16-26 GPa [Fig. 2(a)]. The compressibility of a materials is generally defined as the relative change rate of dimensions with respect to pressure at constant temperature, $K_{i}$ = -(1/$i$)($\partial$$i$/$\partial$$p$)$_{T}$, where $i$ can be assigned as $V$, $A$, and $l$ for volume, area and linear compressibility, respectively \cite{R02}. The compressibility coefficients of the principal axes (equivalent to crystallographic axes in orthorhombic phase) for $Pbam$ phase, computed by the PASCal program in the pressure range from 15.8 to 26.3 GPa \cite{R42}, are $K_{a}$ = -6.65 TPa$^{-1}$, $K_{b}$ = 12.10 TPa$^{-1}$ and $K_{c}$ = 1.94 TPa$^{-1}$, respectively. The variation of the compressibility of the principal axes as a function of pressure is plotted in Fig. 2(b), which shows NLC along the $a$-axis while positive line compressibility (PLC) along the $b$-, $c$-axis. The mechanism of NLC can be rationalized by its special geometrical configuration.

In the unit cell of $Pbam$ phase, the distance $r$ and angle $\theta$ with respect to three nearest centers of neighboring B$_{6}$ octahedra are illustrated in Fig. 2(c). The lattice constants, $a$ and $b$, have the relationship with the parameters of $r$ and $\theta$ as follows:
a = 2$r\cos(\theta$/2), b = 2$r\sin(\theta$/2);
$K_{a}$ = -(1/$a$)($\partial$$a$/$\partial$$p$) = $K_{r}$ + $K_{\cos(\theta/2)}$,
$K_{b}$ = -(1/$b$)($\partial$$b$/$\partial$$p$) = $K_{r}$ + $K_{\sin(\theta/2)}$.
It is evident that the values of $r$ and $\theta$ ($\theta$ $\textless$ 90$\degree$) decrease as pressure increases from 15.8 to 26.3 GPa (Fig. S4) \cite{R35}, resulting in a positive $K_{r}$, a positive $K_{\sin(\theta/2)}$ and a negative $K_{\cos(\theta/2)}$. The pressure-dependent variations of $K_{r}$, $K_{\sin(\theta/2)}$ and $K_{\cos(\theta/2)}$ are depicted in Fig. 2(d). Because $K_{r}$ is much smaller than $K_{\sin(\theta/2)}$ and $K_{\cos(\theta/2)}$, the compressibility of the $a$-axis ($b$-axis) is dominantly determined by the corresponding $K_{\cos(\theta/2)}$ ($K_{\sin(\theta/2)}$) in $Pbam$ phase. Compared with the compression behavior of $P$4/$mbm$ phase, the pressure-induced opposite rotation of B$_{6}$ octahedra in $Pbam$ phase significantly enhances the change rate of $\theta$ above 16 GPa, which guarantees a small $K_{r}$ and a large $K_{\theta}$ value, resulting in NLC accordingly.

\begin{figure}[t]
\begin{center}
\includegraphics[width=7.0cm]{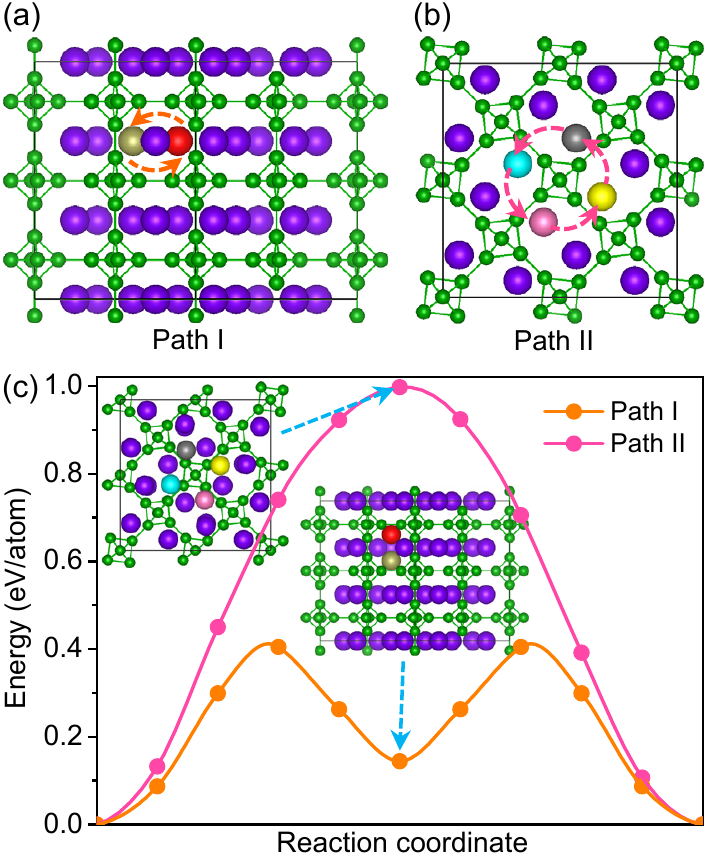}
\caption{%
(a,b) Potential Na diffusion paths, where local exchanges for the Na pairs (path I) and inter-paired Na atoms (path II). The color balls represent the exchanged Na atoms.
(c) Calculated energy barriers of $P$4/$mbm$ phase with different Na diffusion paths. The insets show the structural illustrations of the saddle point at path I and path II.}
\end{center}
\end{figure}

\begin{figure}[t]
\begin{center}
\includegraphics[width=7.0cm]{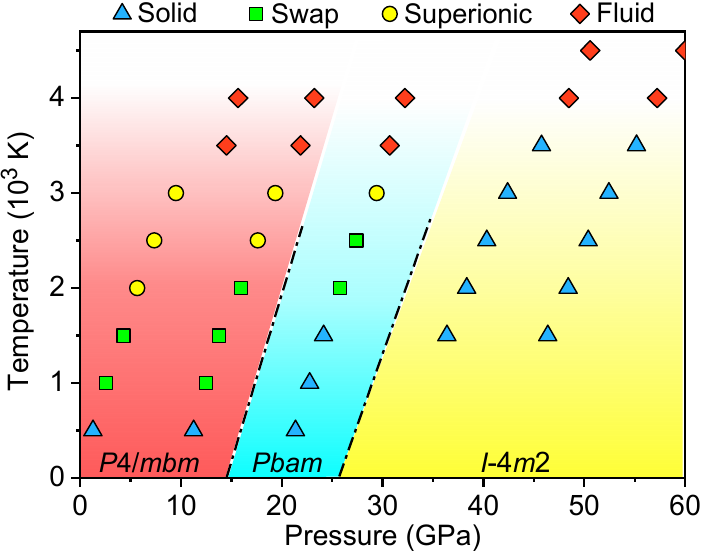}
\caption{%
The phase diagram of NaB$_{3}$ system. The symbols of blue triangles, green squares, yellow circles and jacinth diamonds represent solid, swap, superionic and fluid states, respectively. The data points marked by color symbols in red, cyan and yellow shaded areas relate to $P$4/$mbm$, $Pbam$ and $I$-$4m2$ phases, respectively.}
\end{center}
\end{figure}

To further examine the thermodynamic stability of $P$4/$mbm$ phase, we performed AIMD simulations at ambient pressure and the temperature range of 500-4500 K by analyzing the combination of mean square displacements (MSDs) and atomic trajectories. At 500 K, the oscillations of Na and B atoms with respect to their equilibrium positions without any migrations indicate that the structure is stable, hence the $P$4/$mbm$ structure maintains the solid phase [Figs. 3(a) and 3(d)]. At 1500 K, the MSDs of Na atoms demonstrate step-like increase in several picoseconds, indicating a special displacement during this stage [Fig. 3(b)]. Upon a more detailed study of the atomic trajectories, the local exchange among Na atoms is observed and designated as an intermediate state, as shown in Fig. 3(e) and Supplemental Material \cite{R35}. The Na pairs composed of the nearest neighboring Na atoms rotate $\sim$90$\degree$ nearly parallel to the (110) or ($\bar{1}$10) plane, and then rotate to the initial positions or exchanged positions. This isolated rotation behavior is similar to that of molecular crystals, such as low-temperature phase of H$_{2}$ in hexagonal-close-packed (hcp) form \cite{R43}. However, unlike H$_{2}$ molecules, there is a strong electrostatic repulsion between Na cations instead of covalent bonds. Therefore, the intermediate state of NaB$_{3}$ may be termed as a `swap' phase, which will be further discussed in the later section. At 3000 K, the Na atoms vibrate strongly and migrate freely within a fixed open-framework boron structure, resulting in a persistently ascendant MSDs with a diffusion constant of approximately 0.69 $\times$ 10$^{-8}$ m$^{2}$s$^{-1}$. All these indicate $P$4/$mbm$ phase turns into a superionic conductor [Figs. 3(c) and 3(f)]. When the temperature is increased beyond 3500 K, $P$4/$mbm$ phase melts.

Figures 3(g)-3(i) show the probability density maps of Na atoms \cite{R44}, indicating the diffusive feature of Na atoms in $P$4/$mbm$ phase. The coordinates were first built, with the center \textbf{r}$_{c}$ = ($\bm{\mu}_{i}$(0) + $\bm{\mu}_{j}$(0))/2 = $\bm{\mu}_{i}(t)$ - $\bm{\mu}^{'}_{i}(t)$ defined as the mid-point of the Na-Na vector, where $i$, $j$ represents the atomic number of the nearest neighboring Na atoms, and \bm{$\mu$}$_{i}(t)$ represents the position vector of the $i$ th atoms at the moment of $t$. We defined $\hat{\bm{\eta}}$ as the nearest Na-Na vector and the coordinate $\eta$ = $\bm{\mu}^{'}_{i}(t)\cdot\hat{\bm{\eta}}$ of a given atom $i$. Coordinate $\rho$ is defined as $\rho$ =$\sqrt{\verb||||\bm{\mu}^{'}_{i}(t)\verb||||^{2} - \eta^{2}}$ = $\verb||||\bm{\mu}^{'}_{i}(t)$ - $\eta\hat{\bm{\eta}}\verb||||$. At 500 K, the $P$4/$mbm$ phase retains the solid state [Fig. 3(g)]. When temperature further increases to 1500 K, the diffusive trajectories of Na atoms involving the rotation behavior of Na pairs can be characterized by a diamond shape with two diagonal lines of $\sim$2.43 {\AA} (the distance of the Na pairs in the initial position) and $\sim$2.39 {\AA} (the distance of Na pairs rotating $\sim$90$\degree$ relative to its initial positions) [Fig. 3(h)]. All Na pairs have similar distance of approximately 2.14$\sim$2.43 {\AA} with fixed mass centers, implying a swap state. To reveal the likely mechanism of this particular state, the potential migration pathway of Na atoms is studied by the CI-NEB method, where the initial and final states are identical but with different diffusion paths [Figs. 4(a) and 4(b)]. The results show the energy barrier of the swap state (0.39 eV/atom) is lower than that of inter-paired Na atoms (0.99 eV/atom) [Fig. 4(c)], indicating that the asymmetric energy barrier may be dominantly responsible for the origin of the swap state. When temperature reaches 3000 K, Na atoms diffuse freely within the fixed boron frameworks, entering a superionic state [Fig. 3(i)]. According to the multiple transition states discussed above, we expanded the pressure range up to 60 GPa to explore the phase diagram of NaB$_{3}$. The systematical AIMD simulations at HPHT show that the $Pbam$ phase also exhibits a solid - swap - superionic - fluid phase transition similar to the $P$4/$mbm$ phase (Fig. S5) \cite{R35}, while the $I$-$4m2$ phase transforms directly from solid to fluid above 4000 K [Fig. S6 and Fig. 5] \cite{R35}.

\section{CONCLUSION}
In summary, the phase diagram and compression behavior of NaB$_{3}$ have been systematically studied from first principles. Under high pressure, a solid-solid phase transition originated from the symmetry-breaking of boron framework results in a striking NLC effect. At high temperature, a phase transition sequence of solid - swap - superionic - fluid state is identified by the AIMD simulations. Interestingly, different from the normal solid-superionic phase transition, the particular swap state attributed to the local diffusion of Na pairs may give a deeper insight into the understanding of ion transportation.

\section*{ACKNOWLEDGMENTS}
This work was supported by the National Science Foundation of China (Grants 11874224, 52025026, 52090020, 51772263, and 21803033), the Tianjin Science Foundation for Distinguished Young Scholars (Grant No. 17JCJQJC44400) and Young Elite Scientists Sponsorship Program by Tianjin (No. TJSQNTJ-2018-18). X. D. and X.-F. Z. thank the computing resources of Tianhe II and the support of Chinese National Supercomputer Center in Guangzhou.



\end{document}